\documentstyle[twoside]{article}

\catcode`\@=11
\long\def\@makefntext#1{
\protect\noindent \hbox to 3.2pt {\hskip-.9pt  
$^{{\eightrm\@thefnmark}}$\hfil}#1\hfill}		

\def\thefootnote{\fnsymbol{footnote}}
\def\@makefnmark{\hbox to 0pt{$^{\@thefnmark}$\hss}}	
	
\def\ps@myheadings{\let\@mkboth\@gobbletwo
\def\@oddhead{\hbox{}
\rightmark\hfil\eightrm\thepage}   
\def\@oddfoot{}\def\@evenhead{\eightrm\thepage\hfil
\leftmark\hbox{}}\def\@evenfoot{}
\def\sectionmark##1{}\def\subsectionmark##1{}}

\oddsidemargin=\evensidemargin
\renewcommand{\thefootnote}{\fnsymbol{footnote}}

\newcounter{sectionc}\newcounter{subsectionc}\newcounter{subsubsectionc}
\renewcommand{\section}[1] {\vspace{12pt}\addtocounter{sectionc}{1} 
\setcounter{subsectionc}{0}\setcounter{subsubsectionc}{0}\noindent 
	{\tenbf\thesectionc. #1}\par\vspace{5pt}}
\renewcommand{\subsection}[1] {\vspace{12pt}\addtocounter{subsectionc}{1} 
	\setcounter{subsubsectionc}{0}\noindent 
	{\bf\thesectionc.\thesubsectionc. {\kern1pt \bfit #1}}\par\vspace{5pt}}
\renewcommand{\subsubsection}[1] {\vspace{12pt}\addtocounter{subsubsectionc}{1}
	\noindent{\tenrm\thesectionc.\thesubsectionc.\thesubsubsectionc.
	{\kern1pt \tenit #1}}\par\vspace{5pt}}
\newcommand{\nonumsection}[1] {\vspace{12pt}\noindent{\tenbf #1}
	\par\vspace{5pt}}

\newcounter{appendixc}
\newcounter{subappendixc}[appendixc]
\newcounter{subsubappendixc}[subappendixc]
\renewcommand{\thesubappendixc}{\Alph{appendixc}.\arabic{subappendixc}}
\renewcommand{\thesubsubappendixc}
	{\Alph{appendixc}.\arabic{subappendixc}.\arabic{subsubappendixc}}

\renewcommand{\appendix}[1] {\vspace{12pt}
        \refstepcounter{appendixc}
        \setcounter{figure}{0}
        \setcounter{table}{0}
        \setcounter{lemma}{0}
        \setcounter{theorem}{0}
        \setcounter{corollary}{0}
        \setcounter{definition}{0}
        \setcounter{equation}{0}
        \renewcommand{\thefigure}{\Alph{appendixc}.\arabic{figure}}
        \renewcommand{\thetable}{\Alph{appendixc}.\arabic{table}}
        \renewcommand{\theappendixc}{\Alph{appendixc}}
        \renewcommand{\thelemma}{\Alph{appendixc}.\arabic{lemma}}
        \renewcommand{\thetheorem}{\Alph{appendixc}.\arabic{theorem}}
        \renewcommand{\thedefinition}{\Alph{appendixc}.\arabic{definition}}
        \renewcommand{\thecorollary}{\Alph{appendixc}.\arabic{corollary}}
        \renewcommand{\theequation}{\Alph{appendixc}.\arabic{equation}}
        \noindent{\tenbf Appendix \theappendixc #1}\par\vspace{5pt}}
\newcommand{\subappendix}[1] {\vspace{12pt}
        \refstepcounter{subappendixc}
        \noindent{\bf Appendix \thesubappendixc. {\kern1pt \bfit #1}}
	\par\vspace{5pt}}
\newcommand{\subsubappendix}[1] {\vspace{12pt}
        \refstepcounter{subsubappendixc}
        \noindent{\rm Appendix \thesubsubappendixc. {\kern1pt \tenit #1}}
	\par\vspace{5pt}}

\newcommand{\textlineskip}{\baselineskip=13pt}
\newcommand{\smalllineskip}{\baselineskip=10pt}

\def\eightcirc{
\begin{picture}(0,0)
\put(4.4,1.8){\circle{6.5}}
\end{picture}}
\def\eightcopyright{\eightcirc\kern2.7pt\hbox{\eightrm c}} 

\newcommand{\copyrightheading}[1]
	{\vspace*{-2.5cm}\smalllineskip{\flushleft
	{\footnotesize International Journal of Modern Physics A, #1}\\
	{\footnotesize $\eightcopyright$\, World Scientific Publishing
	 Company}\\
	 }}

\def\abstracts#1#2#3{{
	\centering{\begin{minipage}{4.5in}\baselineskip=10pt\footnotesize
	\parindent=0pt #1\par 
	\parindent=15pt #2\par
	\parindent=15pt #3
	\end{minipage}}\par}} 

\newcommand{\bibit}{\nineit}

\renewenvironment{thebibliography}[1]
	{\frenchspacing
	 \ninerm\baselineskip=11pt
	 \begin{list}{\arabic{enumi}.}
	{\usecounter{enumi}\setlength{\parsep}{0pt}
	 \setlength{\leftmargin 12.7pt}{\rightmargin 0pt} 
	 \setlength{\itemsep}{0pt} \settowidth
	{\labelwidth}{#1.}\sloppy}}{\end{list}}

\newcounter{itemlistc}
\newcounter{romanlistc}
\newcounter{alphlistc}
\newcounter{arabiclistc}

\newcommand{\fcaption}[1]{
        \refstepcounter{figure}
        \setbox\@tempboxa = \hbox{\footnotesize Fig.~\thefigure. #1}
        \ifdim \wd\@tempboxa > 5in
           {\begin{center}
        \parbox{5in}{\footnotesize\smalllineskip Fig.~\thefigure. #1}
            \end{center}}
        \else
             {\begin{center}
             {\footnotesize Fig.~\thefigure. #1}
              \end{center}}
        \fi}

\newcommand{\tcaption}[1]{
        \refstepcounter{table}
        \setbox\@tempboxa = \hbox{\footnotesize Table~\thetable. #1}
        \ifdim \wd\@tempboxa > 5in
           {\begin{center}
        \parbox{5in}{\footnotesize\smalllineskip Table~\thetable. #1}
            \end{center}}
        \else
             {\begin{center}
             {\footnotesize Table~\thetable. #1}
              \end{center}}
        \fi}

\def\@citex[#1]#2{\if@filesw\immediate\write\@auxout
	{\string\citation{#2}}\fi
\def\@citea{}\@cite{\@for\@citeb:=#2\do
	{\@citea\def\@citea{,}\@ifundefined
	{b@\@citeb}{{\bf ?}\@warning
	{Citation `\@citeb' on page \thepage \space undefined}}
	{\csname b@\@citeb\endcsname}}}{#1}}

\newif\if@cghi
\def\cite{\@cghitrue\@ifnextchar [{\@tempswatrue
	\@citex}{\@tempswafalse\@citex[]}}
\def\citelow{\@cghifalse\@ifnextchar [{\@tempswatrue
	\@citex}{\@tempswafalse\@citex[]}}
\def\@cite#1#2{{$\null^{#1}$\if@tempswa\typeout
	{IJCGA warning: optional citation argument 
	ignored: `#2'} \fi}}

\def\pmb#1{\setbox0=\hbox{#1}
	\kern-.025em\copy0\kern-\wd0
	\kern.05em\copy0\kern-\wd0
	\kern-.025em\raise.0433em\box0}

\def\fnt#1#2{\footnotetext{\kern-.3em
	{$^{\mbox{\scriptsize #1}}$}{#2}}}

\def\fpage#1{\begingroup
\voffset=.3in
\thispagestyle{empty}\begin{table}[b]\centerline{\footnotesize #1}
	\end{table}\endgroup}

\def\runninghead#1#2{\pagestyle{myheadings}
\markboth{{\protect\footnotesize\it{\quad #1}}\hfill}
{\hfill{\protect\footnotesize\it{#2\quad}}}}
\font\tenrm=cmr10
\font\tenit=cmti10 
\font\tenbf=cmbx10
\font\bfit=cmbxti10 at 10pt
\font\ninerm=cmr9
\font\nineit=cmti9

\font\eightrm=cmr8

\def\str{\mbox{str}}
\def\sdet{\mbox{sdet}}
\newcommand{\mcal}[1]{{\mathcal #1}}

\def\qed{\hbox{${\vcenter{\vbox{			
   \hrule height 0.4pt\hbox{\vrule width 0.4pt height 6pt
   \kern5pt\vrule width 0.4pt}\hrule height 0.4pt}}}$}}

\renewcommand{\thefootnote}{\fnsymbol{footnote}}	

\begin{document}

\runninghead{Physical results from partially quenched simulations}
{Physical results from partially quenched simulations}

\normalsize\textlineskip
\thispagestyle{empty}
\setcounter{page}{1}

\copyrightheading{}			

\vspace*{0.88truein}

\fpage{1}
\centerline{\bf PHYSICAL RESULTS from PARTIALLY QUENCHED
SIMULATIONS\footnote{Based on talks given by S. Sharpe and N. Shoresh
at DPF 2000, August 2000.}}
\vspace*{0.37truein}
\centerline{\footnotesize STEPHEN R. SHARPE and NOAM SHORESH}
\vspace*{0.015truein}
\centerline{\footnotesize\it 
Physics Department, University of Washington, Box 351560}
\baselineskip=10pt
\centerline{\footnotesize\it Seattle, Washington 98195-1560, USA}

\vspace*{0.21truein}
\abstracts{We describe how one can use chiral perturbation theory
to obtain results for physical quantities, such as
quark masses, using partially quenched simulations.}{}{}

\textlineskip			
\vspace*{12pt}			

\setcounter{footnote}{0}
\renewcommand{\thefootnote}{\alph{footnote}}

\vspace*{-0.5pt}
\noindent

For some time to come, simulations of lattice QCD will
not work directly with physical up and down quarks.
This is because the computer time required
scales roughly as $m_q^{-2.5}$ with existing algorithms
(at fixed physical box size).
Present simulations, using machines sustaining up to $0.3$ Teraflops,
are limited to quark masses greater than about half the
strange quark mass, an order of magnitude larger than the average
up and down quark mass.
Even a dedicated machine sustaining $10$ Teraflops
(hopefully to be available in 2003-4) will allow 
quark masses to be reduced to only $\sim m_s/8$.

Thus an extrapolation in light quark masses is required.
Fortunately, this can be done, using chiral perturbation theory (ChPT),
once the simulated masses are small enough.
In practice, it is feasible to determine the functional forms
with which to extrapolate at next-to-leading order
(NLO) in the chiral expansion.
A useful way of thinking about the extrapolation is that
simulations with moderately light quark masses can be used to
determine the parameters of the chiral Lagrangian (including the
Gasser-Leutwyler coefficients $L_{1-10}$ which appear at NLO),
and then the extrapolation can be done ``by hand''.

In this talk we describe some recent work in which we show how
the chiral extrapolation can be aided by the use of partially quenched
(PQ) simulations\cite{SSlat99,SS,SSnew}. 
These are simulations in which the ``valence'' and
``sea'' ($=$ ``dynamical'') quarks have different masses.
The key theoretical observation is that, 
if both valence and sea quarks are light enough,
then the chiral Lagrangian describing the long distance
properties of the PQ simulations contains the same parameters 
($f$, $\langle \bar q q\rangle$, and $L_{1-10}$) as
appear in the chiral Lagrangian for QCD~\cite{SPQ}.
This follows from the work of Ref.~\cite{BGPQ}.
Thus one can extrapolate to the physical theory using PQ as well
as unquenched simulations, without introducing new, unphysical, parameters.
This is true despite the fact that the PQ theory itself is unphysical.%
\footnote{A clear example of the unphysical nature of PQ theories is
that some flavor-singlet correlation functions have double poles.}

The practical importance of this observation has yet to be seen,
but could be significant. It is relatively cheap (CPU $\sim m_V^{-1}$)
to reduce the valence quark masses at fixed sea quark mass,
and, in fact, most calculations including sea quarks do undertake
extensive PQ simulations. Our point (amplified in examples given
below) is that these simulations should not be viewed as
a somewhat improved quenched approximation, from which one can only
obtain qualitative information, but rather as a quantitative
technical tool that one can use to obtain physical parameters.

Note that it is essential for the PQ simulations to be done with the
same number of light dynamical quarks as in physical QCD,
since the parameters of the chiral Lagrangian depend on this number.
This brings up the tricky question of whether the strange quark is
light enough to be described accurately enough by ChPT at NLO.
The answer depends on the physical quantities being considered.
For quantities involving the light pseudoscalar mesons, ChPT
including the strange quark seems to be reasonably convergent
[NLO corrections are generically $\sim M_K^2/(4 \pi f_\pi)^2$].
Thus it is probably appropriate to work in the sector with three
light quarks.
For baryonic quantities, where the expansion parameter is
$M_K/(4 \pi f_\pi)$, the situation is much less clear.
Here one may be forced to consider a two-flavor chiral Lagrangian
with $m_s$-dependent parameters, in order to make use of PQ simulations.
In any case, the beauty of lattice simulations is that
they can be used, in principle,
to resolve these questions by comparing numerical
results to the different theoretical predictions.

We now turn to a more detailed explication of these general remarks.
This will be done using the properties of pseudo-Goldstone mesons
(henceforth referred to generically as ``pions'') as an example.
More details can be found in Refs.~\cite{SS,SSnew}.

We begin by recalling the form of the chiral Lagrangian
for PQQCD~\cite{BGPQ},
\begin{eqnarray}
\mcal{L}&=&\frac{f^2}{4}
\str \left( \partial_\mu U\partial_\mu U^\dagger\right)
-\frac{f^2}{4}\str\left(\chi U^\dag+U\chi\right)
-L_6\left[\str\left(\chi U^\dag+U\chi\right)\right]^2 + \dots,
\label{eq:PQChL}
\end{eqnarray}
where, for brevity we have displayed only one of the NLO terms.
Note that $\mcal{L}$ has the same form as that for QCD,
except that $U=\exp(2i\Pi/f)$ is an element of the graded
group $SU(5|2)$ rather than $SU(3)$
(assuming three sea quarks and two valence quarks),
that traces are replaced by supertraces ($\str$),
and that the mass matrix which appears in $\chi= 2 \mu M$
(with $\mu = -\langle\bar q q\rangle/f^2$) is enlarged to
\begin{equation}
M = \mbox{ diag}(m_A,m_B,m_u,m_d,m_s,m_A,m_B)
\end{equation}
with $m_{A,B}$ being the masses of the valence quarks and the
corresponding ghosts.\footnote{%
In the subsequent expressions,
these masses can be taken to be lattice quark masses,
defined through Ward Identities, up to corrections of $O(a)$.
See Ref.~\cite{SS} for more discussion.} 

An important theoretical point is that the ``pion'' field $\Pi$
is ``straceless'', $\str\Pi=0$, so that $U$ is an element of $SU(5|2)$
rather than $U(5|2)$. Since the symmetry breaking pattern is
$SU(5|2)_L\times SU(5|2)_R \to SU(5|2)_V$, this means that 
there is one element of $\Pi$ for each Goldstone particle.
This all sounds very reasonable and hardly worthy of note.
We mention it for two reasons.
First, because it is in sharp contrast to the situation in the fully
quenched theory. The quenched chiral Lagrangian
must be constructed using a field which is an element of
the unitary graded group, e.g. $\Sigma\in U(2|2)$ for two valence quarks.
The non-anomalous chiral symmetries then allow arbitrary 
functions of $\Phi_0 \propto \ln\sdet \Sigma$ in the Lagrangian
(as long as they are consistent with parity), including the
well-known mass term $m_0^2 \Phi_0^2$.
The theory is then no longer predictive,
because there are an infinite number of new couplings, 
and no power counting scheme with which to order them.
One must proceed in a phenomenological manner, assuming that 
most couplings are small, and ignoring higher-order $\Phi_0$ loops.
The essence of the problem is that the physics due to the
scale $m_0\sim 1\;$GeV cannot be integrated out.

The second reason we raise this point is that, if you look at papers
doing calculations in PQChPT, including our own, you will find
that $U$ is enlarged to $\Sigma=U\exp(i 2\Phi_0/\sqrt3 f)\in U(5|2)$.
Given the discussion in the previous paragraph,
you might well wonder why one would include the $\Phi_0$ and 
lose predictive power.
In our case, it was because we did not realize until recently
that one could work without the $\Phi_0$, and so we followed a
similar path to that for the quenched theory.
The good news, however, is that one can show that if one takes previous
results obtained including the $\Phi_0$, 
and sends $m_0^2\to\infty$, one recovers the
results that would have been obtained if $\Phi_0=0$ all along~\cite{SSnew}.
In fact, it is technically easier to do the calculations this way,
rather than to explicitly project against $\Phi_0$.

In summary, for light enough sea quark masses
one can non-perturbatively integrate out the $\Phi_0$ from the PQ theory.
This result was anticipated in Refs.~\cite{SPQ,GLPQ}
on the basis of a one-loop analysis.

As our first application,
we have calculated the pion masses and decay constants for PQ theories
with an arbitrary number of sea quarks with arbitrary masses~\cite{SS},
extending the work of Refs.~\cite{SPQ,GLPQ} for degenerate sea quarks.
This allows a determination of $f$, $\mu$ and $L_{4-8}$.
We find, for example, that the mass of the flavor non-singlet meson created
by the operator $\bar A\gamma_5 B$ is
\begin{equation}
M_{AB}^2 = 
\overline\chi_V \left[ 1 + {8N\over f^2} (2 L_6\!-\!L_4) \overline\chi_S
+ {8N\over f^2} (2 L_8\!-\!L_5) \overline\chi_V + \mbox{\rm chiral logs}
+O(p^4) \right]\,,
\label{eq:MAB}
\end{equation}
where $\overline\chi_{V,S}$ 
are the average valence and sea quark masses, respectively,
multiplied by $2 \mu$.
The ``chiral logs'' are generically of the form $c(\chi/f^2) \ln\chi$,
with $c$ a known number, and so do not introduce any new constants.
The dependence on the scale in the logarithm is cancelled by
the implicit scale dependence of the Gasser-Leutwyler coefficients.
Obtaining these logarithms for arbitrary quark masses
is the hard part of the calculation.
Note that Eq.~(\ref{eq:MAB}) includes results for the unquenched theory;
these can be obtained by setting $m_A$ and $m_B$ equal to sea quark masses.

Fitting Eq.~(\ref{eq:MAB}) to numerical results
(assuming $f$ is known from fits to the decay constants)
one can, in principle, determine $2 L_6\!-\!L_4$ and $2 L_8\!-\!L_5$.
With unquenched simulations alone, however, a separate
determination of these two combinations requires
non-degenerate sea quarks
(for otherwise $\overline\chi_V=\overline\chi_S$).
By contrast, with PQ simulations a separate determination is
straightforward since $\overline\chi_S$ and $\overline\chi_V$ are 
independent, and one need only use degenerate sea quarks. 
It is particular noteworthy that the combination
$2 L_8\!-\!L_5$, which is related to the value of the physical up-quark mass
as discussed below,
can be determined from the dependence on $\overline\chi_V$, 
using only a {\em single} mass for the degenerate sea quarks~\cite{CKN}.
Of course, this mass must be light
enough that NNLO corrections are small.

This discussion shows how the extra ``dials'' that one can adjust
in the PQ theory simplify the extrapolation to unquenched QCD.
We stress, however, that we are not particularly advocating the use of 
three {\em degenerate} sea quarks---the so-called ``$2+1$'' simulations
with $m_u=m_d < m_s$ work just as well, and might be preferred
as they approach closer to the physical parameters.
Our general formulae for the chiral logs apply to either case.

It seems that it will be practical to carry out
such determinations in the next few years, 
and we spend the rest of this talk
discussing issues that might arise.
First recall that we only have good experimental information
on two combinations of $L_{4-8}$ (all values are quoted at the scale
$M_\eta$):
\begin{equation}
L_5=(2.3\pm 0.2) \times 10^{-3}\,,\qquad
2 L_7+L_8= (0.4\pm 0.1) \times 10^{-3} \,.
\label{eq:Ls}
\end{equation}
The remaining 3 linear combinations will either be
very difficult, or impossible~\cite{KaplanManohar}, 
to learn from experiment alone. One can estimate these 3 combinations
using phenomenological models, and there are a fairly standard
set of values that can be found in texts and review articles.
Thus, determining $L_{4-8}$ will allow both a test of QCD
(Do these physical parameters equal the experimental values of 
Eq.~(\ref{eq:Ls})?)
and a test of phenomenological models.

Perhaps the most interesting aspect of any such determination
is the light it might shed on the value of $m_u$,
the physical up-quark mass.\footnote{This point has been stressed
in Ref.~\cite{CKN}, where more details of the following discussion
can be found.}
A vanishing value solves the strong-CP problem.
At NLO in ChPT, and taking the experimental meson masses as inputs,
the value of $m_u$ depends on the combination $2 L_8\!-\!L_5$. 
The standard parameters (which correspond to $m_u/m_d \approx 1/2$) 
give $2 L_8\!-\!L_5\approx 0$,
while to obtain $m_u=0$ one needs a value
in the range $(-1.2)-(-2.6) \times 10^{-3}$.
There is a large uncertainty since, in this scenario,
NLO corrections are $\approx 50\%$, and so NNLO uncertainties are large.
The upper and lower ends of this range come 
from writing the NLO expressions either in
terms of quark masses or of squared meson masses.
Because of this uncertainty, one can only hope to
rule out $m_u=0$ by determining the $L_i$.
To demonstrate that $m_u=0$ will be much more difficult.

We have examined how results from PQ simulations would look
in these two scenarios: the ``Standard L's'' or ``$m_u=0$ L's''.\footnote{%
Some results for the former appear in Ref.~\cite{SS}, while those for
the latter are new.}
\ For simplicity, we set $m_u=m_d$,
and consider only charged meson properties, so $L_7$ does not enter.
We set $L_4=L_6=0$ in both scenarios, and\footnote{%
For the $m_u=0$ scenario,
$L_5$ is about half the value quoted in Eq.~(\ref{eq:Ls}) because we
write the NLO expressions in terms of quark masses.}
\begin{equation}
(L_5,L_8)\times 10^3 \approx
(2.3,\, 1.2)\  [\mbox{standard}]\quad \mbox{or}\quad
(1.2,\, 0) \ [m_u=0] 
\end{equation}
Our values of $f$ and the quark masses
are then chosen to reproduce the physical ratios 
$M_\pi/M_K$, $f_\pi/f_K$, and $f_\pi/M_\pi$, within a small tolerance.
We stress that our choices of parameters are not unique---indeed, there is not
enough experimental information to fix all the parameters---
but rather that they are representative of the two scenarios.

Meson properties depend on four parameters: the two valence quark masses
and the light and strange sea quark masses.
To reduce this to two parameters,
we show results for ``kaon-like'' mesons,
those in which the masses one of the valence quarks and the strange sea
quark are fixed to the physical strange quark mass ($m_{\rm st}$).
The properties then depend on the light valence and sea quark masses.
If both are equal to the physical average light quark mass then
our kaon is very close to the physical kaon.
The quantities we plot are the ratios of NLO to LO contributions:
\begin{equation}
\delta_K^M = (M_K^2/\bar\chi_V) - 1 \,. \quad
\mbox{and}\quad
\delta_K^f = (f_K /f ) - 1 \,. \quad
\end{equation}
These should have magnitudes small compared to unity
for the chiral expansion to be reliable.
The kaon-like mesons
include some examples of the poorest convergence of the chiral expansion.

\begin{figure}[tb]
\vspace*{2.8 in}
\includegraphics{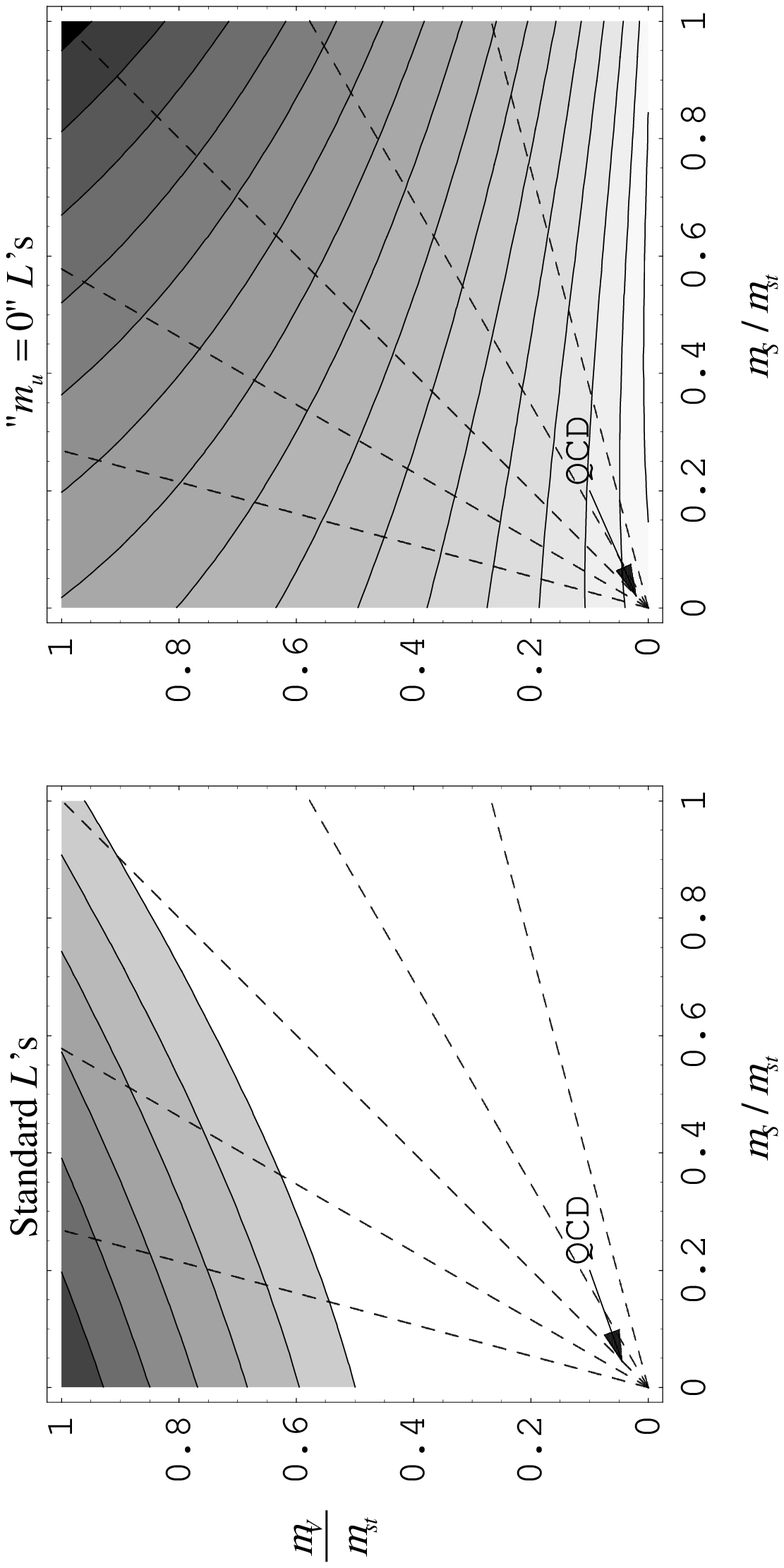}
\vspace*{-0.2 in}
\fcaption{Contour plots of $\delta_K^M$. The x- and y-axes are the light
sea and valence quark masses, respectively,
in units of the physical strange quark mass.
In the left-hand plot, the white region has $0<\delta_K^M<0.1$,
and the contours increase from $0.1$ to $0.2$ (upper left) in steps of $0.02$.
In the right-hand plot the contours start at $0.475$ (lower left) 
and increase to $0.825$ in steps of $0.025$.}
\label{fig:contour}
\end{figure}

In Fig.~\ref{fig:contour} we show contour plots of $\delta_K^M$ 
for the two sets of parameters. 
Unquenched simulations lie on the diagonal,
while most PQ simulations lie below the diagonal.
Most present simulations lie to right of the $x=0.5$ line. 
We observe that the chiral expansion is well
behaved for standard parameters
(and, indeed, the PQ region below the diagonal has better convergence
than the unquenched line), while the expansion is not reliable
for the $m_u=0$ case. 
All one can say in the latter case is that there
should be substantially more curvature in the plots of $M_K^2$ versus
quark mass than for standard parameters.

\begin{figure}[tb]
\vspace*{2.6 in}
\includegraphics{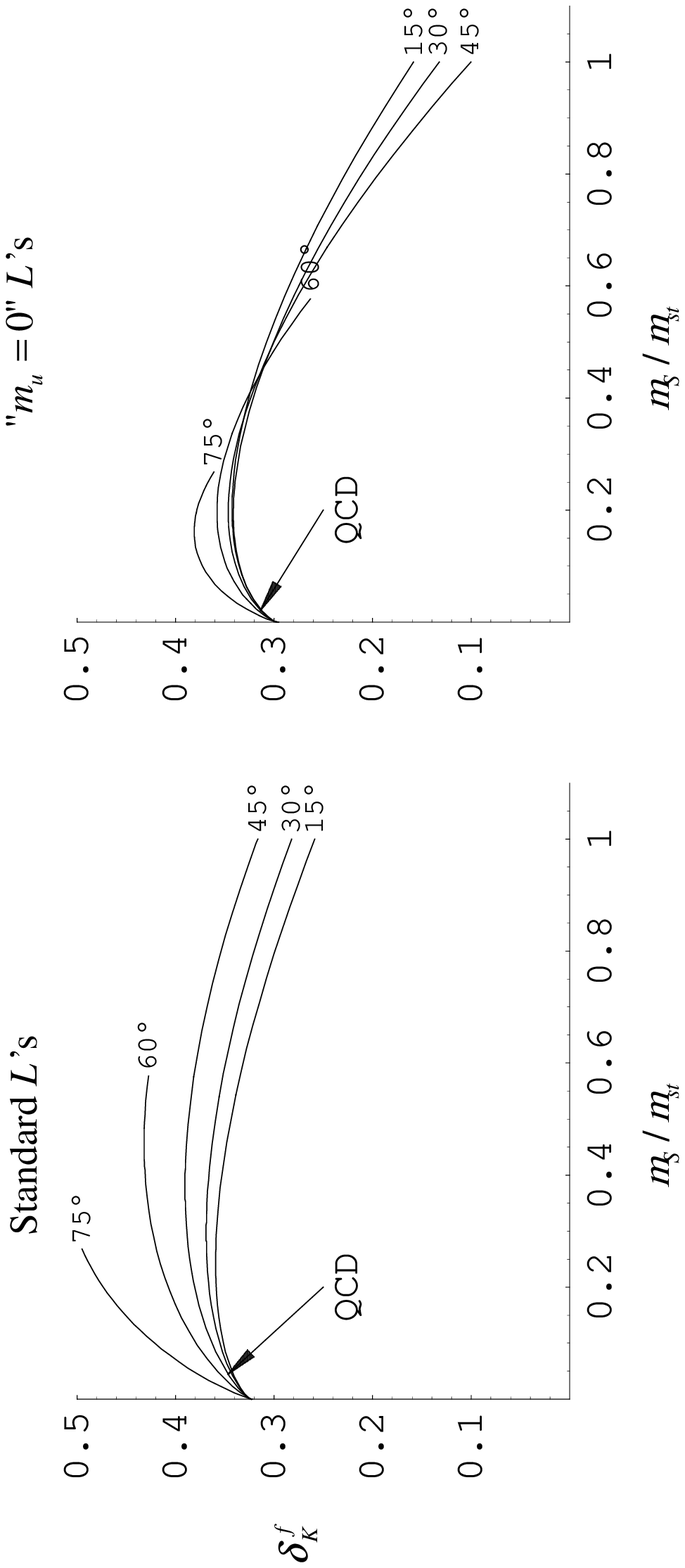}
\vspace*{-0.2 in}
\fcaption{$\delta_K^f$ along rays of fixed angle in the contour plots.}
\label{fig:rays}
\end{figure}

In Fig.~\ref{fig:rays} we plot the NLO/LO ratio for
decay constants  of kaon-like mesons. 
We plot slices through the contour plot of
$\delta_K^f$ (not itself shown)
along the various ``rays'' 
shown by the dashed lines in Fig.~\ref{fig:contour},
with the angle being that w.r.t. the x-axis.
In this case the chiral expansion is reasonably well behaved for
both sets of parameters, although the results themselves differ substantially.

These plots bring out a warning to those fitting PQ or unquenched
data. If one were to fit the curves for, say, $15^\circ-45^\circ$,
in the region $0.5< x < 1$ with a straight line and extrapolate,
one would make a $10-20\%$ error in the final result for $f_K$.
This is exactly the fit which would result if
one kept only the analytic terms,
and dropped the chiral logs,
in expressions such as Eq.~(\ref{eq:MAB}).
Clearly one must keep the logarithms when extrapolating. 
We stress that this does not not introduce new parameters,
and that this is true not only for PQ extrapolations
but also for extrapolations along the unquenched line at $45^\circ$.

\nonumsection{Acknowledgements}
\noindent
This work was supported in part by U.S. Department of Energy
grant No. DE-FG03-96ER40956/A006.

\nonumsection{References}

\end{document}